\documentclass{iau} 
\usepackage{graphicx}
\usepackage{natbib}
\def\msun{$M_{\odot}$~}
\def\aap{\it A\&A \rm}
\def\apj{\it ApJ \rm}

\def\nat{\it Nature \rm}
\def\mnras{\it MNRAS\rm}
\def\an{\it AN \rm}

\title[The oldest stars of the bulge] 
{The oldest stars of the bulge: new information on the ancient Galaxy}

\author[G. Cescutti, C. Chiappini \& R. Hirschi]   
{Gabriele Cescutti$^1$,
Cristina Chiappini$^2$
\and Raphael Hirschi$^{3,4}$}

\affiliation{$^1$I.N.A.F. Osservatorio Astronomico di Trieste, Italy; 
\\ email: {\tt cescutti@oats.inaf.it} \\[\affilskip]
$^2$Leibniz-Institut für Astrophysik Potsdam (AIP) \\[\affilskip]
$^3$ Astrophysics Group, Faculty of Natural Sciences, Keele University \\[\affilskip]
$^4$ Kavli IPMU (WPI), University of Tokyo}

\pubyear{2017}
\volume{334}  
\jname{Rediscovering our Galaxy}
\editors{C. Chiappini,
I. Minchev,
E. Starkenburg \&
M. Valentini eds.}
\begin{document}

\maketitle

\begin{abstract}
  Recently the search for the oldest stars have started to focus on
  the Bulge region. The Galactic bulge hosts extremely old stars, with
  ages compatible with the ages of the oldest halo stars. The data
  coming from these recent observations present new chemical
  signatures and therefore provide complementary constraints to those
  already found in the halo.  So, the study of the oldest bulge stars can improve
  dramatically the constraints on the nature of first stars and how
  they polluted the pristine ISM of our Galaxy. We present our first
  results regarding the light elements (CNO) and the neutron capture
  elements. Our findings in the oldest bulge stars support the
  scenario where the first stellar generations have been fast
  rotators.

\end{abstract}
\section{Introduction}
In our recent work, we have provided an interpretation of the presence
in halo stars of specific chemical signatures by means of stochastic
chemical evolution models
\citep[][]{Chiappini13}. Our results supported
the scenario in which the first stars that exploded and polluted the
pristine interstellar medium (ISM) were rotating faster than the
present day massive stars. Stellar evolution codes coupled with
nuclear reaction chains have shown that this rotation produces mixing
in the interior of the stars. This mixing impacts the nucleosynthesis
of light elements such as carbon, nitrogen, and oxygen, and it also
predicts the production of s-process elements
\citep{Frischknecht16}. In this scenario in which the stars were fast
rotating, chemical evolution models were able to explain several
chemical anomalies observed in the early Universe: the almost solar
ratio of [N/O] and the increase and spread in the [C/O] ratio
\citep{Chiappini06}; the low $^{12}$C/ $^{13}$C ratios
\citep{Chiappini08}; the spread present in the [C/O] and [N/O] ratios
\citep{Cesc10} and between light and heavy neutron capture elements
\citep{Cescutti13,Cescutti14,Cescutti15,Cescutti16}.

The same nucleosynthesis acting in other environments produces
different results due to the different star formation (SF) histories.
Therefore it is important to test our predictions in other
environments that have hosted  the explosions of the
first stars.  In our neighbourhood, we expect to find signatures
of the enrichment  by the first stars also in the stars belonging to the dwarf
spheroidal galaxies and the oldest stars of the Galactic bulge.  In
this proceedings, we focus on the \emph{old bulge}, that is
seen as a sort of an extreme case of the halo population, with an
increased density and faster chemical enrichment.
The Galactic bulge is a very complex component of our Galaxy
 and it is likely that different populations share this locus, with
 different histories of SF and chemical enrichment.
 Recent investigations have highlighted the presence of a very old
 component and the evidence shows that the efficiency of 
 chemical enrichment has been higher in this system compared to the
 Galactic halo. A higher efficiency can be quite naturally connected
 to a higher density on the ISM. This has been also
 supported by recent evidence of the existence of mature bulges at
 redshifts of z$\sim$2.2 \citep{Tacchella15}, and is also
 predicted by models of strong early turbulent gas accretion \citep[see][and references therein]
 {Bournard16}.

\section{Stochastic chemical evolution model for the Bulge}

\begin{figure*}[ht!]
\begin{center}
\includegraphics[width=100mm]{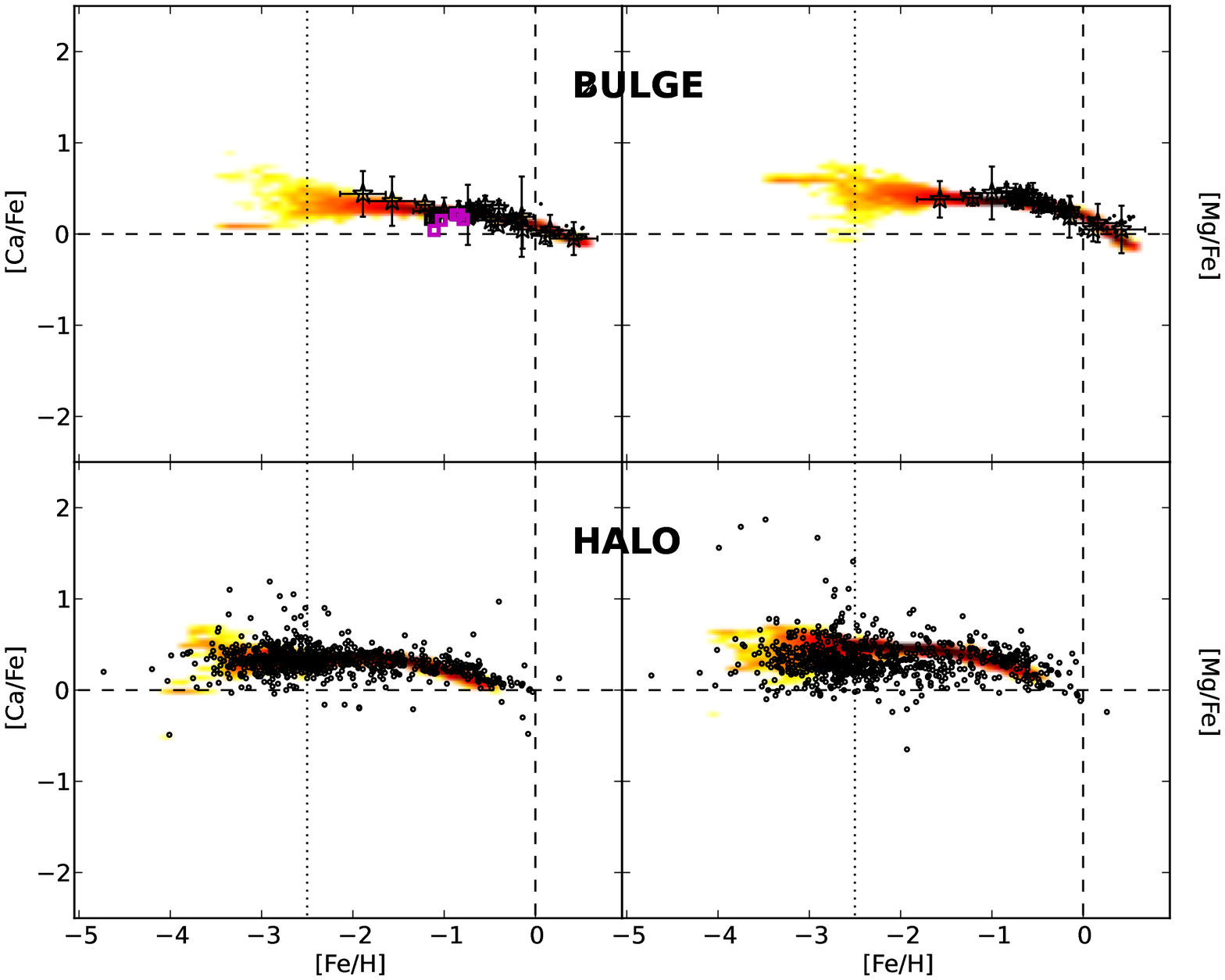}
\includegraphics[width=90mm]{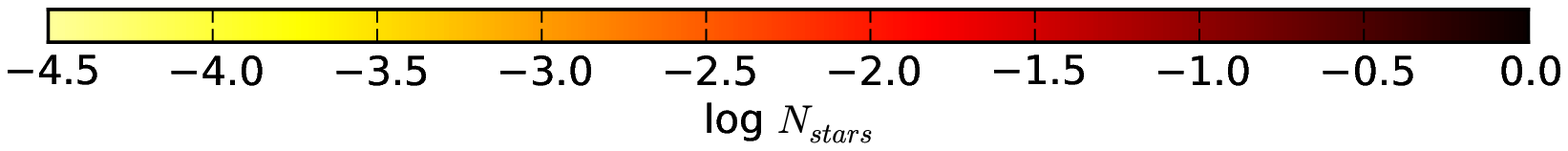}

\caption{[Ca/Fe] and [Mg/Fe] vs [Fe/H], from left to right. We present
  the results of the bulge model in the upper panels and of the halo
  model in the lower panels \citep{Cescutti13}.  The density plots
  are the distribution of simulated long-living stars for our models;
  the density is on a logarithmic scale and the bar under the figure
  describes the assumed color scale.  The abundance ratios for bulge
  stars (upper panels) and halo stars (lower panels) are shown
  \citep[for the data used
  see][]{Cescutti13,Cescutti17a}.}\label{alphas}

\end{center}
\end{figure*}

Our chemical evolution model for the \emph{old bulge} is an extreme
case of the halo model. Compared to the halo model, it has an
increased density ($\sim$ 10x) and SF efficiency
($\sim$20x) with no outflows. The increased efficiency is consistent
with the SF efficiencies adopted in other chemical
evolution models for the bulge \citep{Cescutti11}. The nucleosynthesis
prescriptions are those adopted in \citet{Cescutti14}.
  The bulge model is constrained by reproducing the metallicity distribution of
the oldest bulge stars, as well as by the [$\alpha$/Fe], see Fig. \ref{alphas}.
On this classical chemical evolution scheme, we
implement a stochastic formation of stars similar to
the chemical evolution model for the halo \citep{Cesc08}.  Compared to
the halo model the length scale for
the mixing zone has been decreased from 90pc (for the halo) to 30pc
(for the bulge).  The modification of this typical length scale is due
to the dependency of the dimension of the SNe bubble to the ISM
density. More details are available in \citet{Cescutti17a}.

\section{Results adopting two different r-process events}

\begin{figure*}[ht!]
\hspace{-1.cm}
\includegraphics[width=70mm]{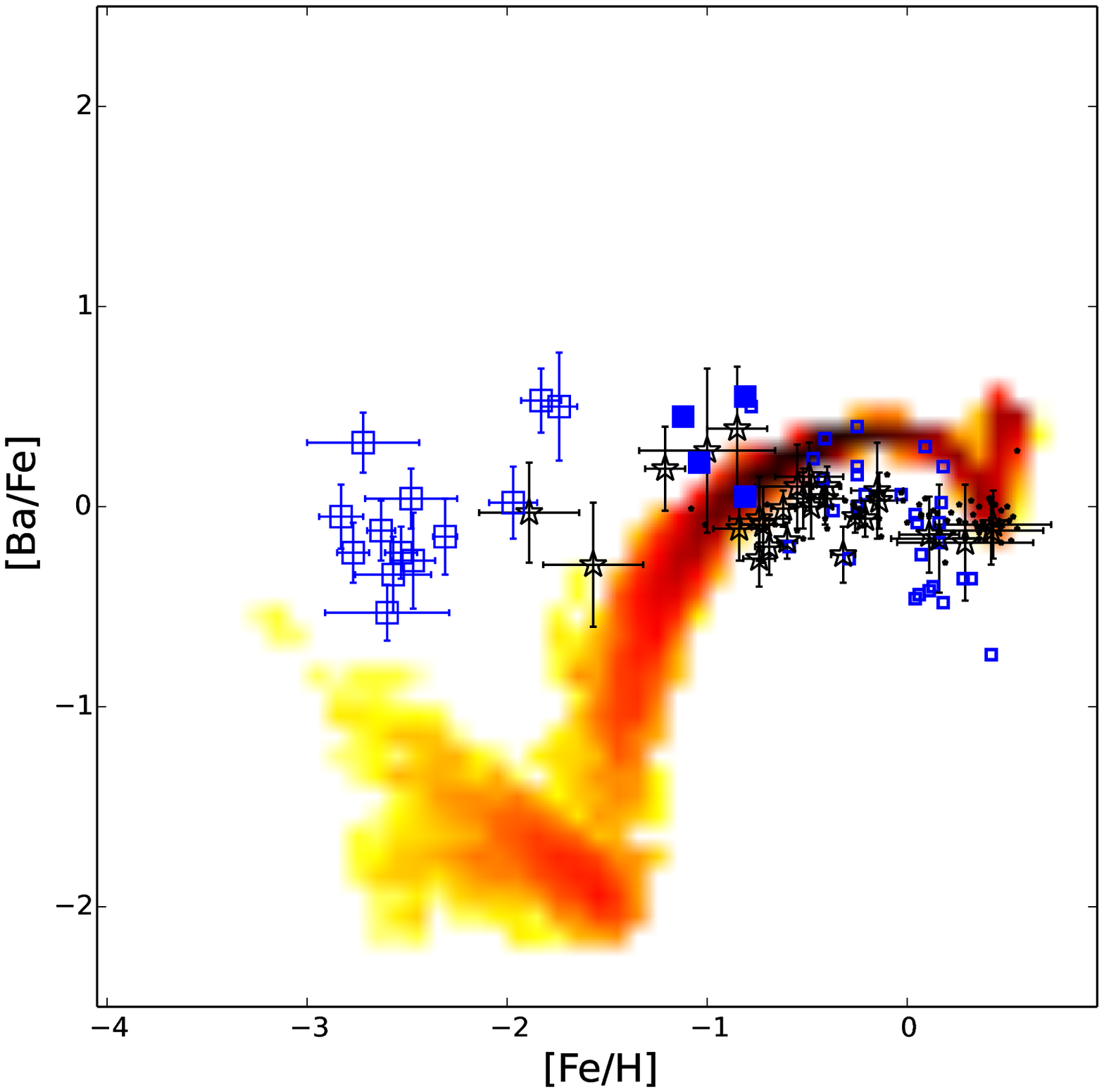}
\hspace{-0.5cm}
\includegraphics[width=70mm]{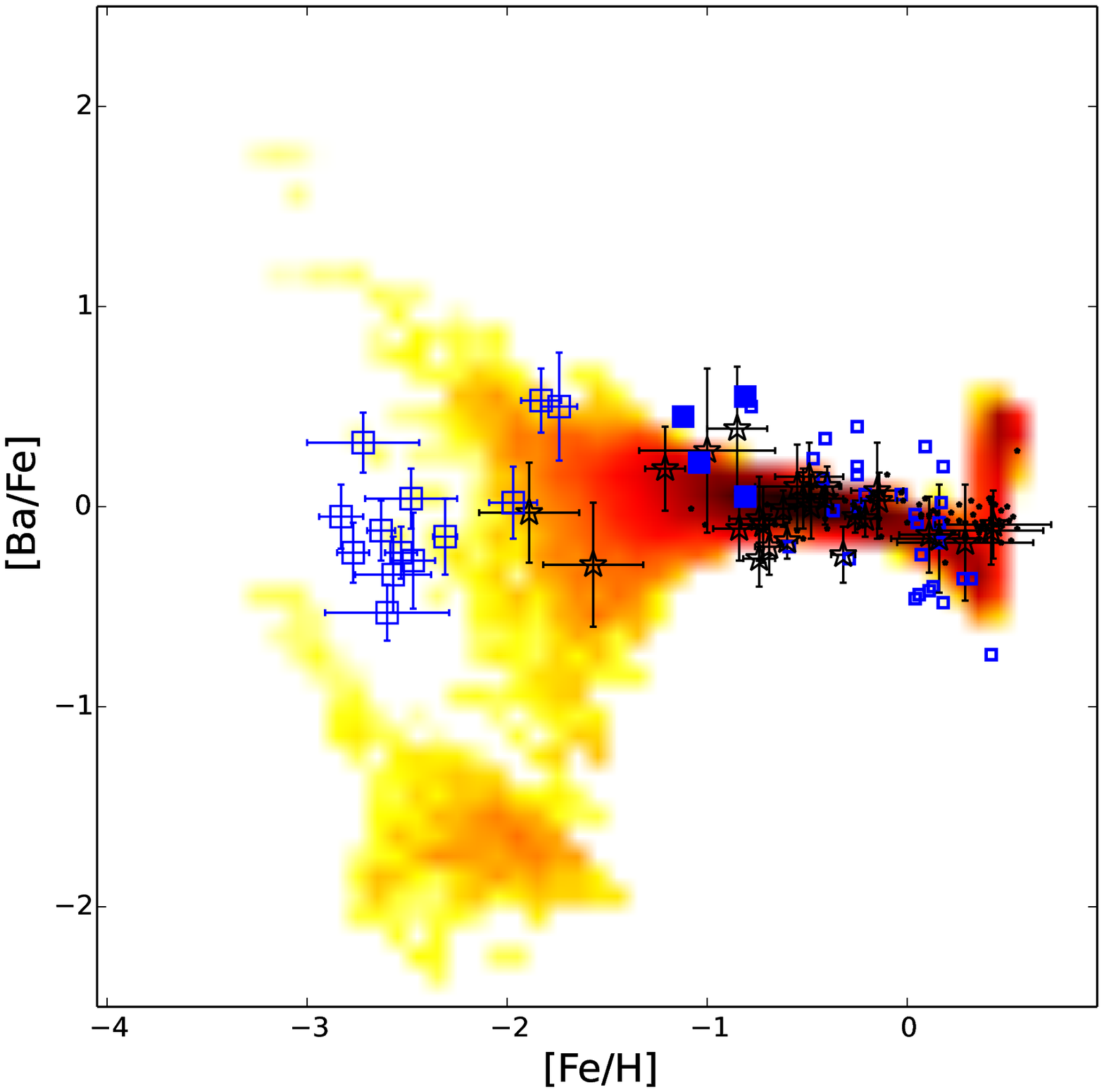}

\caption{[Ba/Fe] vs [Fe/H].  In the figures above we present the
  results of the bulge model; on the left with the
  nucleosynthesis \emph{EC+s2}, on the right with the one \emph{MRD +s
    B2}. In both figures we show abundance ratios for bulge stars
  \citep[for details of the different authors
  see][]{Cescutti17b}}\label{Bafig}

\end{figure*}

We display the bulge results adopting the same nucleosynthesis
prescriptions adopted in the halo model results shown in
\citet{Cescutti14}, where the details of the different nucleosynthesis
can be found.  We display only the results for [Ba/Fe] as a typical
case of neutron capture element.  In the left panel of
Fig.\ref{Bafig}, the same nucleosynthesis of the \emph{EC+s2} model of
\citet{Cescutti14} is adopted for the the bulge model.  We
briefly recall that in this model an r-process contribution is assumed
from all the stars between 8-10 \msun, assumed to explode as
electron capture supernovae (EC SNe).

In this model, due the intense SF there after 20-30 Myr
(the lifetime of a 8-10 \msun star) all the individual volumes have already
exploded massive stars despite of the stochasticity; this was not the
case for the halo model. So, all the r-process events turn out to
happen in an environment which is already enriched in Fe and not so
inhomogeneous; for this reason the r-process in this model does not
produce an important spread but simply a rise on the [Ba/Fe] vs
[Fe/H] space moving from low metallicity to solar metallicity. The
spread at lower metallicity is driven by the differential enrichment
by the spinstars as a function of their masses.
The stellar distribution obtained does not appear to be consistent to
the data.  In the bulge the delay between the enrichment by r-process
and the spinstars can be appreciated and it produces the late rise
of the [Ba/Fe] ratio at around [Fe/H]$\sim -$1.5, whereas the Ba
enrichment up to this metallicity is manly due to s-process by
spinstars.  The combination of the two different productions is the
cause of the ``V'' shape visible in this plot.
Focusing on the most metal poor stars in the sample,
the observational data present a higher [Ba/Fe] ratio compared to the
model results.  This is a signature that within the timescale of the
chemical evolution of the bulge the EC scenario is not well
supported as r-process event.

In the right panel of Fig.\ref{Bafig}, we show the results assuming
the nucleosynthesis model named \emph{MRD +s B2} in
\citet{Cescutti14}.  We recall that in this model r-process events are
assumed from 10\% of all the massive stars that explode as magneto
rotationally driven supernovae \citep[MRD SNe, see][]{Winteler12}.
With these nucleosynthesis prescriptions the bulge model produces
results completely different to the EC+s model; in particular the
\emph{MRD +s B2} model is able to explain much better the data in the
bulge stars for [Ba/Fe], whereas for the halo the results were very
similar \citep[see][]{Cescutti14}.  The different model predictions
for the halo and bulge are due to the vigorous and fast chemical
evolution in the bulge model coupled to the differences between this
two possible r-process events, and the difference chiefly resides in
the timescales for the two sites of production, which varies from 30
Myr to less than 5 Myr \citep[for more details see][]{Cescutti17b}.

\section{Conclusions}

We apply the nucleosynthesis for two different r-process events in our
stochastic modelling for the old bulge. The nucleosynthesis resulting
from either of these r-process were able to reproduce the chemical enrichment in
[Ba/Fe] observed in halo stars.  The comparison of the results
obtained for the bulge model to the [Ba/Fe] abundances measured in the
bulge supports a r-process event with a time scale very close to the
lifetime of massive stars, as the MRD scenario and it seems to exclude
r-process events with longer timescale, as the EC scenario.  This
outcome is also compatible with neutron star mergers as r-process
events, providing short merging timescales \citep{Cescutti15}.

\begin{acknowledgements}
  
  G.C. acknowledges financial support from the European Union Horizon
  2020 research and innovation programme under the Marie Sk\l
  odowska-Curie grant agreement No. 664931.  
RH acknowledges support from the World Premier
  International Research Center Initiative, MEXT, Japan.
C. C. acknowledges support from DFG Grant CH1188/2-1. This work has been
  partially supported by the the EU COST Action CA16117 (ChETEC) and
  by the European Research Council (EU-FP7-ERC-2012-St Grant
  306901). 

\end{acknowledgements}


\end{document}